\documentstyle[psfig,aps,prl,amssymb]{revtex}

\begin{document}
\title{Exact Bures Probabilities that Two Quantum Bits are
Classically Correlated}
\author{Paul B. Slater}
\address{ISBER, University of California, Santa Barbara, CA 93106-2150\\
e-mail: slater@itp.ucsb.edu, FAX: (805) 893-7995}

\date{\today}

\draft

\maketitle

\vskip -0.1cm

\begin{abstract}
In  previous studies, we have explored the {\it ansatz} that the volume
elements of the {\it Bures}  metrics over 
quantum systems
might serve  as {\it prior} distributions, in analogy with the 
(classical) {\it Bayesian} role of the volume elements (``Jeffreys' priors'')
of {\it Fisher information} metrics.
Continuing this work, we obtain  {\it exact} Bures prior 
probabilities that the members of 
certain {\it low}-dimensional subsets of the {\it fifteen}-dimensional 
convex set of $4 \times 4$ 
density matrices are {\it separable}
 or {\it classically
correlated}.
The  main analytical tools employed are {\it symbolic} integration and 
a formula of Dittmann (J. Phys. A 32, 2663 [1999]) 
for Bures metric 
tensors.
This study complements an earlier one (J. Phys. A 32, 
5261 [1999]) in which numerical
(randomization) --- but {\it not} integration --- methods
 were used to estimate Bures separability probabilities for 
{\it unrestricted} $4 \times 4$ and $6 \times 6$ density matrices.
The exact values adduced here for pairs of quantum bits (qubits), 
typically, well exceed the estimate ($\approx .1$) there, but
this disparity may be attributable to our
focus on special low-dimensional subsets. 
Quite remarkably, for 
 the $q= 1$ and $q = {1 \over 2}$ states inferred using the
principle of maximum nonadditive (Tsallis) 
entropy, the Bures probabilities 
of separability are {\it both} equal to $\sqrt{2} -1$.
For the Werner {\it qubit-qutrit}
 and {\it qutrit-qutrit}
states, the probabilities 
are vanishingly small, while in the {\it qubit-qubit} case it is
${1 \over 4}$.
\end{abstract}


\vspace{.1cm}

\tableofcontents

\newpage

\section{Introduction}

\subsection{Background}

In a previous study \cite{slatZHSL}, we exploited
certain  {\it numerical} methods to 
estimate the {\it a priori} probability --- based on the volume element 
of the Bures metric 
\cite{ditt1,ditt2,hub1,hub2,braun} --- that, a  member of the 
fifteen-dimensional convex set ($R$) of $4 \times 4$ density matrices 
is {\it separable} (classically correlated), that is, expressible as a
convex combination 
of tensor products of 
pairs of $2 \times 2$ 
density matrices. (Ensembles of separable states, as well as of 
{\it bound entangled} 
states  can not be ``distilled'' to obtain  pairs 
in  singlet states for 
{\it quantum} information processing \cite{HORO3,VEDRAL}.)
This Bures probability estimate $ \approx .1 $ was 
rather unstable in character \cite[Table 1]{slatZHSL},
 due, in part it appeared, to 
difficult-to-avoid ``over-parameterizations'' of $R$, as well as to 
the unavailability, in that context, of numerical {\it integration} methods. 
But now in secs.~\ref{secthreea}, \ref{wow} and \ref{secthreec} below, 
we are able to report {\it exact} probabilities of separability   by 
restricting  
consideration to certain low-dimensional subsets of $R$, for which 
{\it symbolic} integration can be performed. Then, in the subsequent body of
the paper, we investigate analogous questions when $R$ is replaced by the
convex sets $9 \times 9$ and $6 \times 6$ density matrices.
(We have also studied, using numerical methods, the Bures probability of
separability of the two-party {\it Gaussian} states \cite{slateressent} 
(cf. \cite{clifton}).)

Preliminarily though, we
 investigate in sec.~\ref{sectwo} certain relevant 
motivating issues, first having arisen in
 the context of the $3 \times 3$ density
 matrices. These quantum-theoretic
entities belong to an eight-dimensional convex set ($Q$), which 
we parameterize in the manner,
\begin{equation} \label{paramden}
\rho_{Q}  ={1 \over 2} \pmatrix{v +z & u - \mbox{i} w & x - \mbox{i} y \cr
u + \mbox{i} w & 2 -2 v & s - \mbox{i}  t \cr
x + \mbox{i} y & s + \mbox{i} t & v -z \cr} .
\end{equation}
The feasible 
range of the eight parameters --- defined by the boundary of $Q$ --- is
 determined by the requirements imposed on density matrices, in general,
 that they 
be Hermitian, nonnegative definite (all 
eigenvalues nonnegative),
and have unit trace \cite{bloore}.

Dittmann \cite[eq. (3.8)]{ditt1} (cf. \cite{ditt2}) 
has presented an ``explicit'' formula  
(one not requiring the computation of eigenvalues and eigenvectors) for
the Bures metric 
(\cite{ditt1,ditt2,hub1,hub2,braun})
 over the $ 3 \times 3$ density matrices. It takes the form
\begin{equation} \label{ditty1}
d_{Bures}(\rho,\rho + \mbox{d} \rho)^{2}
 = {1 \over 4} \mbox{Tr}
\{ \mbox{d} \rho \mbox{d} \rho + {3 \over  1 -\mbox{Tr} \rho^{3}} (\mbox{d} \rho 
-\rho \mbox{d} \rho) (\mbox{d} \rho -\rho \mbox{d} \rho) 
+{3 |\rho| \over 1 - \mbox{Tr} \rho^{3}} (\mbox{d} \rho -\rho^{-1} 
\mbox{d} \rho) (\mbox{d} \rho -\rho^{-1} \mbox{d} \rho) \}.
\end{equation}
If we implement this formula, using $\rho_{Q}$ for $\rho$, 
we obtain an $8 \times 8$ matrix --- the Bures metric tensor, which we will
denote by $g$.

It has been proposed
 \cite{slatZHSL,slatjpa,slatjpa2,slatjmp1,slatcomp,slathall}
 that the
 square root of the determinant of $g$,
that is, $|g|^{1/2}$, which gives  the {\it volume element} of the
metric, be taken
as a prior
distribution  
(to speak in terms of  the specific instance 
presently before us) over the $3 \times 3$ 
density matrices (cf. \cite{ksieee}). This {\it ansatz} 
is based on an analogy with Bayesian theory 
\cite{kass,bernardo}, in
which the volume element of the {\it Fisher information}
\cite{frieden} matrix is used as a
{\it reparameterization-invariant} prior, termed  ``Jeffreys' prior''.

Unfortunately, the 
(``brute force'') 
computation of the determinant of such $8 \times 8$ 
symbolic matrices appears to  exceed present capabilities 
\cite{murao,krattenthaler}. 
In light of this limitation, we pursued a strategy of
 fixing (in particular, setting to zero) 
a certain number (four) of the eight parameters, thus,  
leading to an achieveable  calculation.
A similar course was followed in a brief exercise in 
\cite[eqs. (31), (32)]{slatjmp1}, but 
using a quite different
parmeterization 
of  $Q$ --- one 
 based on the expected values with respect to
 a set of four mutually unbiased (orthonormal) bases 
of three-dimensional Hilbert space \cite{wootters,wootters2}. 

In 
\cite{slathall} we have reported exact results for the ``Hall normalization 
constants'' for the Bures volumes of the $n$-state quantum systems,
$n =2,\ldots,6$.
These analyses utilized certain
  parameterizations (of Schur form)
 of the $n \times n$ density matrices \cite{boya}, in which 
the eigenvalues and eigenvectors of these 
density matrices are {\it explicitly} given. It was 
established there \cite[sec. II.B]{slathall}, among other things, that the 
Bures volume element for the $3 \times 3$ density matrices is, in fact, 
{\it normalizable} over $Q$, forming a {\it probability distribution}.
 Since it appears to
be highly  problematical to find an explicit transformation from this  
eigen-parameterization of Boya {\it et al} \cite{boya} to that 
used in (\ref{paramden}), we can not 
conveniently utilize the results of \cite{slathall} for our 
specific purposes here. For the $n$-level systems,
$n > 3$, the analogous task would appear to be 
even more challenging, since the appropriate 
parameterizations of $SU(n)$ and the associated invariant (Haar) 
measures seem not yet to have 
been developed (cf. 
\cite{byrd1,byrd2,mallesh,rowe}).

\subsection{{\it Two} forms of conditional Bures priors
 for a {\it four}-parameter {\it three}-level 
system} \label{sectwo}

 Since we aim to reexamine the specific findings in
\cite{slatjpa}, we will thus cast our analyses 
specifically in terms of the
 parameterized form (\ref{paramden}). We have computed, using the formula 
(\ref{ditty1}), the 
$8 \times 8$ Bures metric tensor ($g$) associated with (\ref{paramden}).
Then --- only subsequent to this computation --- we set the
 four parameters $s$, $t$, $u$ and $v$ all 
equal to zero in $g$, obtaining what we denote by 
$\tilde{g}$. (Actually, this ``conditioning'' on $P$ can be performed 
immediately after the 
determination of the differential element $\mbox{d} \rho$, thus simplifying 
the further calculations in (\ref{ditty1}).) 
Then,
\begin{equation} \label{new}
{|\tilde{g}|}^{1 \over 2} = 
{1 \over 64 v (1-v)^{1 \over 2} (v^2 -x^2 -y^2 -z^2)^{1 \over 2}
 (x^2 +y^2 +z^2 - (v-2)^2)},
\end{equation}
which could be considered to constitute the (unnormalized) 
conditional Bures prior over
$P$.

On the other hand, if we  {\it ab initio} nullify the same 
four parameters ($s,t,u,v$) 
in $\rho_{Q}$, we get the family of density matrices,
defined over a four-dimensional convex subset ($P$) of $Q$,
\begin{equation} \label{oldden}
\rho_{P} = {1 \over 2} \pmatrix{v + z & 0 & x - \mbox{i} y \cr
0 & 2 - 2v & 0 \cr
x + \mbox{i} y & 0 & v -z \cr},
\end{equation}
which was the specific object of study in \cite{slatjpa}.

Now, let us describe two ways in which an alternative to 
the presumptive conditional prior (\ref{new}), that is, 
\begin{equation} \label{prev}
{1 \over 16 v (1-v)^{1 \over 2} (v^2 -x^2-y^2-z^2)^{1 \over 2}},
\end{equation}
 has been derived.
We obtain the outcome  (\ref{prev}) if we either: (a)
 employ $\rho_{P}$ directly in Dittmann's formula
(\ref{ditty1}), and generate the corresponding $4 \times 4$ metric tensor
 and compute the square root of its 
determinant (the  procedure followed  in 
\cite{slatjpa});  or (b) 
extract from $\tilde{g}$ the $4 \times 4$ submatrix with rows and columns
associated with (the four {\it non}-nullified parameters) 
$v$, $x$, $y$ and $z$, that is
\begin{equation} \label{wierd}
{1 \over 4 (v^2-x^2-y^2-z^2)} \pmatrix{ {v - x^2-y^2-z^2
 \over 1-v} & -x & -y & -z \cr
-x & {v^2 -y^2 -z^2 \over v} & {x y \over v} & {x z \over v} \cr
-y & {x y \over v} & {v^2 -x^2 -z^2 \over v} & { y z \over v} \cr
-z & {x z \over v} & {y z \over v} & {v^2 -x^2 -y^2 \over v} \cr}
\end{equation}
  and 
 calculate the square 
root of its determinant. The matrix (\ref{wierd}) is {\it not} exactly 
the same as
(14) there --- a result which was presumably also obtained 
by the use of $\rho_{P}$ in 
(\ref{ditty1}. The four 
diagonal entries there are the {\it negatives} of the ones in (\ref{wierd}),
that is, we had there \cite[eq. (14)]{slatjpa}
\begin{equation} \label{wierd2}
{1 \over 4 (v^2-x^2-y^2-z^2)} \pmatrix{ {v - x^2-y^2-z^2
 \over 1-v} & -x & -y & -z \cr
-x & {y^2 +z^2 -v^2 \over v} & {x y \over v} & {x z \over v} \cr
-y & {x y \over v} & {x^2 +z^2 -v^2 \over v} & { y z \over v} \cr
-z & {x z \over v} & {y z \over v} & {x^2 +y^2 -v^2 \over v} \cr}
\end{equation}
In any case, the determinants of these two nonidentical $4 \times 4$ 
matrices are the same, so the substantive conclusions of \cite{slatjpa} 
regarding Bures priors are 
unchanged.  

M. J. W. Hall has pointed out that the result (\ref{new}) is, 
in fact, an eight-dimensional volume element rather that the 
four-dimensional one desired here. In addition, a referee has remarked there
can be ``two different sets of basis one-forms that are used to compute the
volume element. This happens, for example, in $SU(n)$ when one uses
$A^{-1} \mbox{d} A$ as the matrix of left invariant one-forms. There exist
$n^2$ invariant forms in this matrix. One must choose an independent set.
This set is thus not unique''.

It is interesting to compare the form of (\ref{wierd}) with the Bures metric
tensor for the $2 \times 2$ systems \cite[eq. (4)]{slatjpa},
\begin{equation}
{1 \over 4 (1-x^2-y^2-z^2)} \pmatrix{1 -y^2 -z^2 & x y & x z \cr
x y & 1 - x^2 -z^2 & y z \cr
x z & y z & 1-x^2 -y^2},
\end{equation}
obtained by the application of Dittmann's formula \cite[eq. (3.7)]{ditt1}
(cf. (\ref{ditty1})),
\begin{equation}
d_{Bures}(\rho,\rho+\mbox{d} \rho)^{2} = {1 \over 4} \mbox{Tr} 
\{ \mbox{d} \rho \mbox{d} \rho +{1 \over |\rho|} (\mbox{d} \rho - \rho
\mbox{d} \rho) (\mbox{d} \rho - \rho \mbox{d} \rho) \}.
\end{equation}
(Of course, in the limit $v \rightarrow 1$, $\rho_{P}$, in effect, degenerates 
to a two-level system, and it is of interest to keep this in mind in examining
the results presented here. In the opposite limit $v \rightarrow 0$, 
one simply leaves the  domain of quantum considerations.)

We note that (\ref{prev}) differs from (\ref{new}) in that it has 
an additional factor, 
\begin{equation} \label{factor1}
f= {1 \over 4 (x^2+y^2+z^2 - (v-2)^2)}.
\end{equation}
 Since $v$ can be no greater than 1
and $x^2+y^2+z^2$  no greater than $v$ if $\rho_{P}$ is
 to meet the nonnegativity requirements of
 a  density
matrix, $f$ must be {\it negative} over the feasible
range ($P$) of parameters of $\rho_{P}$.
In fact, the square root of the determinant of the ``complementary''
$4 \times 4$ submatrix of $\tilde{g}$ --- the one associated with the 
{\it nullified}
parameters, $s,t,u,w$, rather than $v$, $x$, $y$, $z$  --- is equal to 
$f $.

Now, it is interesting to note  --- transforming the Cartesian coordinates
($x$, $y$, $z$) to spherical ones ($r$, $\theta$, $\phi$) --- that
while the previous result (\ref{prev}) of \cite{slatjpa} 
can be normalized to a (proper) 
probability distribution over $P$,
\begin{equation} \label{proper}
 p( v,r,\theta,\phi) = {3 r^2 \sin{\theta} \over 4 \pi^{2} v (1-v)^{1 \over 2} 
(v^{2} - r^2)^{1 \over 2}},
\end{equation}
the new prior (\ref{new}) is  itself {\it not}
 normalizable over $P$, that is, it
is
{\it improper}.
However, we can 
(partially) integrate (\ref{new}) over the three spherical 
coordinates to obtain the {\it univariate}  marginal over 
the variable $v$,
\begin{equation} \label{uninew}
q(v) = {\pi^{2} \over 64 v} (-1 -{2 \over {(1-v)}^{1 \over 2}} + {1 \over -1 +v}).
\end{equation}
The integral of (\ref{uninew}) over $v \in [0,1]$ diverges, however.
We can  compare (\ref{uninew})  with the univariate marginal
 {\it probability distribution} of (\ref{proper}) \cite[eq. (19)]{slatjpa},
\begin{equation} \label{uniold}
p(v)= {3 v \over 4 (1-v)^{1 \over 2}}.
\end{equation}
(In \cite{slattherm1,slattherm2} $p(v)$ was interpreted as a 
density-of-states or structure function, for thermodynamic purposes, and 
the associated partition function reported.) The behaviors of (\ref{uninew}) 
and (\ref{uniold}) are quite distinct, the latter monotonically increasing 
as $v$ increases, while the [negative] of the former has a minimum at 
$v \approx .618034$.

Let us also observe that  the 
factor $f$, given in (\ref{factor1}),  the added presence of 
which leads to the non-normalizability of 
(\ref{new}), 
takes the form in the spherical coordinates,
\begin{equation}
f={1 \over 4 (r+2 -v) (r+v-2)} = {1 \over 4 (r^2 - (v-2)^2)}.
\end{equation}
The eight eigenvalues ($\lambda$) 
of  the nullified form of the Bures metric 
tensor $g$, that is $\tilde{g}$, come in pairs. They are
\begin{equation}
\lambda_{1,2} = {1 \over 4 v}, \qquad 
\lambda_{3,4} = {1 \over 4 + 2 r - 2 v}, \quad
\lambda_{5,6} = -{1 \over 2 (r+v-2)}, 
\end{equation}
\begin{displaymath}
\lambda_{7,8} =
{1 \over -2 \lgroup r^2 + (v-2) v) +2 (r^4 +v^4 + 2 r^2 (2 + (v-4) v) \rgroup^{1 \over 2}}.
\end{displaymath}
Of course, the product of these eight eigenvalues gives us $|\tilde{g}|$, the 
square root of which --- that is (\ref{new}) --- constitutes the
 new  (but unnormalizable/improper) 
possibility here for the conditional Bures/quantum Jeffreys'
prior over the four-dimensional convex subset $P$ of the eight-dimensional
convex set $Q$ composed of the $3 \times 3$ density matrices.

\section{Bures priors and  separability probabilities 
for various composite quantum 
systems} \label{secthree}
\subsection{{\it One}-parameter $2 \otimes 2$ systems} 
\label{secthreea}

Now, let us seek to extend the 
comparative form  of analysis in sec.~\ref{sectwo} to the  
$4 \times 4$ density matrices. For the Bures metric in this 
setting we rely
upon Proposition 1 in the recent paper of Dittmann
\cite{ditt2}, which presents an explicit formula in
terms of the characteristic polynomials of the density matrices.
(Let us point out that in the earlier 
preprint versions, in particular quant-ph/9911058v4, of our
 paper here,
a number of ``anomalous'' results were reported. These turned out to be
attributable to our misinterpretation of the symbol $Y'$ in \cite{ditt2} as
the transpose of $Y$, rather than the conjugate transpose of $Y$. We have
since amended our analyses in this regard.)
We  apply it to several  {\it one}-dimensional 
convex subsets of the fifteen-dimensional convex set ($R$) of $4 \times 4$ 
density matrices. These subsets --- unless otherwise 
indicated ---  are (partially) characterized by 
having their associated two $2 \times 2$ reduced
systems described by the fully mixed (diagonal) density matrix, having
${1 \over 2}$ for its two diagonal entries. Or to put it equivalently, the
three Stokes/Bloch parameters for each of the two subsystems 
are all zero.
(A complete characterization of the inseparable $2 \otimes 2$ systems with
maximally disordered subsystems has been presented within the Hilbert-Schmidt
space formalism \cite{refhoro}.)
\subsubsection{The three intra-directional correlations are
all equal}

For our first scenario, we stipulate
 {\it zero} correlation 
between the spins of these two reduced (fully mixed) systems in {\it different}
 directions, but 
identical non-zero (in general) 
correlation between them in the {\it same} ($x, y$ or $z$) 
 directions. We denote this common correlation
parameter by $\zeta$. In terms of the parameterization of the coupled 
two-level 
systems given in \cite{mk1} (cf. \cite{fano,aravind}), the feasible range of $\zeta$ is
$[-{1 \over 4},{1 \over 12}]$. (The parameterization in \cite{mk1} is based 
on the superposition of sixteen $4 \times 4$ matrices --- which are the
pairwise  direct 
products of the four $2 \times 2$  Pauli matrices, including 
among them, the identity matrix. 
Since the 
six Stokes/Bloch parameters have all been set to
 zero, the nine correlation parameters 
($\zeta_{ij}, i,j = x,y,z$) must all
lie between -1 and 1, and the nine-fold sum of their squares can not exceed 3 
\cite{mk1}. It has been shown that all the tangent vectors corresponding 
to a basis of the Lie algebra --- corresponding to two copies 
of $SU(2)$ --- span six dimensions, and thus there are, in
fact, nine {\it nonlocal} parameters \cite{linden}. A referee has suggested
that the use of ``local orbits would simplify the picture especially for
physicists dealing professionally with entanglement. It is because then  the
reader knows that one deals with what is of the main importance [orbit 
parameters --- like Schmidt coefficients for pure states] from the point of
view of say quantum information transmission [like e. g. teleportation]'' 
(cf. \cite{makhlin,kz}))

\begin{figure}
\centerline{\psfig{figure=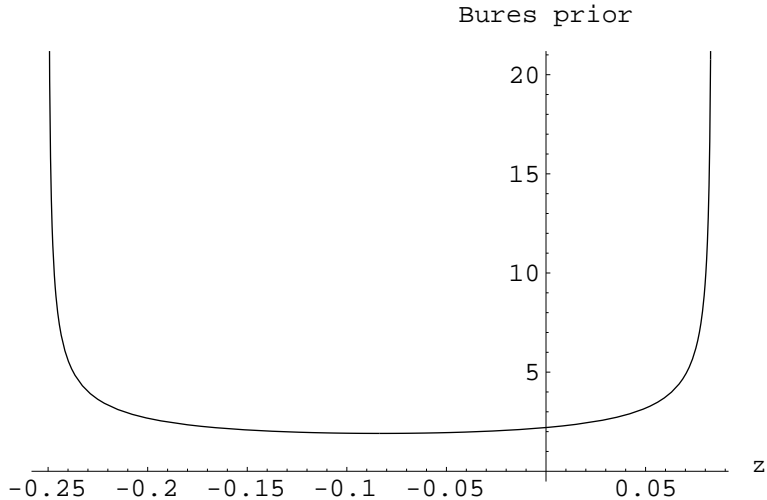}}
\caption{Normalized conditional Bures prior (\ref{works}) for 
one-parameter four-level 
scenario 1}
\label{figfirst}
\end{figure}

If we implement the formula of Dittmann \cite[eq. (9)]{ditt2} 
using a general
(fifteen-parameter) $4 \times 4$ 
density matrix \cite{mk1}, then nullify twelve of the
parameters of the resultant Bures metric tensor, and set
the  indicated remaining three ($\zeta_{i i}$) all equal to one value $\zeta$,
 we obtain as the conditional Bures prior  (the counterpart of 
$|\tilde{g}|^{1 \over 2}$ in sec.~\ref{sectwo}),
\begin{equation} \label{pair1}
{32768   \over (1- 4 \zeta)^{3} (1+ 4 \zeta)^{9 \over 2} 
\sqrt{1 - 12 \zeta}}.
\end{equation}
On the other hand, if we set the fifteen parameters in 
precisely this same fashion {\it before}
 employing the formula of Dittmann,
we obtain for the volume element
\begin{equation} \label{works}
{2 \sqrt{3} \over (1 -8 \zeta -48 \zeta^{2})^{1 \over 2}}.
\end{equation}
The former prior 
is {\it non-normalizable} over $\zeta \in [-{1 \over 4}, {1 \over 12}]$,
while the latter is {\it normalizable}, its 
integral over this interval equalling
${\pi \over 2}$. In Fig.~\ref{figfirst}, we display this probability 
distribution. 
The pair of outcomes ((\ref{pair1}) and (\ref{works})) 
is, thus,  fully analogous 
in terms of normalizability, to what we found above ((\ref{new}) and
 (\ref{prev}))
for the  particular  four-dimensional case ($P$) 
of the three-level quantum systems ($Q$) investigated above.

Now, for $\zeta \in [-{1 \over 12},{1 \over 12}]$, the associated one-parameter
density matrix is {\it separable} or {\it classically correlated} (a 
necessary and {\it sufficient} condition for which for the $4 \times 4$ and
$6 \times 6$ density matrices is that their
 partial transposes have nonnegative eigenvalues \cite{horo3}). So,
if we integrate the normalized form of (\ref{works}) over this interval, 
we obtain the conditional Bures probability of separability
 (cf. \cite{slatZHSL,ZHSL,zyczpra,slateressent}).
 This probability turns out to be precisely 
 ${1 \over 2}$. Contrastingly, in \cite{slatZHSL},
 for {\it arbitrary}  coupled two-level systems in the fifteen-dimensional 
convex set $R$, 
it was necessary to rely upon numerical 
(randomization) simulations for estimates of the 
Bures probability of separability, so this {\it exact} result appears 
quite novel in nature.
(In \cite{slatZHSL}, the [unconditional] Bures probability of separability was
estimated to be $\approx .1$.)

\subsubsection{One intra-directional correlation equals the negative of 
the other two}
 \label{scenario2}

\begin{figure}
\centerline{\psfig{figure=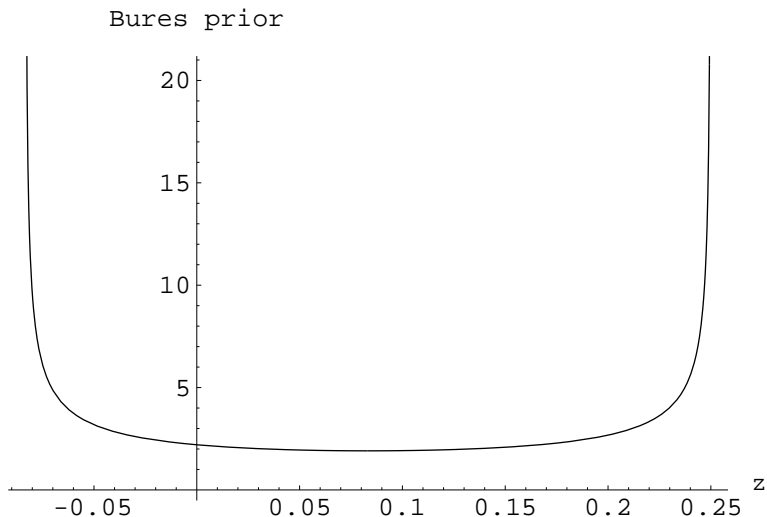}}
\caption{Normalized conditional Bures prior (\ref{SCENE2}) for one-parameter
four-level scenario 2}
\label{scene2}
\end{figure}

A closely related  scenario in which the probability of separability is 
also precisely
${1 \over 2}$ is one for which the only non-nullified parameters are again
the three 
intra-directional correlations, but now two of them 
(say, for the $x$ and $y$-directions) are set equal to $\zeta$
and the third to $-\zeta$. Then, the conditional Bures probability distribution
(computed in the analogous manner) is (Fig.~\ref{scene2})
\begin{equation} \label{SCENE2}
{4 \sqrt{3} \over \pi (1 + 8  \zeta - 48 \zeta^{2})^{1 \over 2}}.
\end{equation}
The region of feasibility is $[-{1 \over 12}, {1 \over 4}]$ and of
separability, $[-{1 \over 12},{1 \over 12}]$.

\subsubsection{The six inter-directional correlations are all equal}

\begin{figure}
\centerline{\psfig{figure=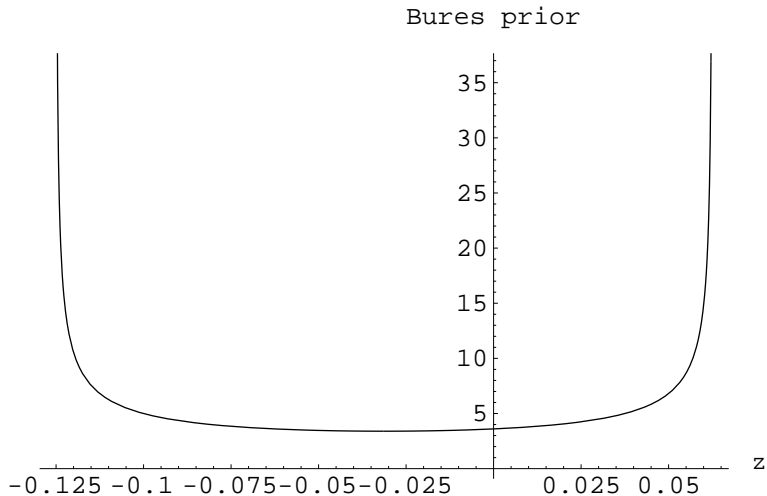}}
\caption{Normalized conditional Bures prior (\ref{SCENE3}) for one-parameter 
four-level scenario 3}
\label{scenario3}
\end{figure}

Let us now examine another 
one-parameter  scenario in which the pair of two-level systems 
is still composed of fully mixed states, but for  which the correlations 
$(\zeta_{i i}$) in 
the same directions are zero, while
 the correlations in different directions ($\zeta_{i j}, i \neq j$) are
not necessarily zero and all equal.  Thus, we 
{\it ab initio} set the (six) interdirectional correlations to $\zeta$,  
the other nine parameters all  to zero, and employ the formula of Dittmann 
\cite{ditt2} (in the manner, we have settled upon for this and all 
subsequent analyses here).
We obtain the conditional Bures {\it probability distribution}
 (Fig.~\ref{scenario3}),
\begin{equation} \label{SCENE3}
{8 \sqrt{2} \over \pi (1-8 \zeta - 128 \zeta^{2})^{1 \over 2}},
\end{equation}
over the feasible range, $\zeta \in [-{1 \over 8},{1 \over 16}]$. 
The range of separability is $[-{1 \over 16},{1 \over 16}]$. The
associated 
conditional Bures probability of separability is then 
${1 \over 2} + {\sin^{-1}{1 \over 3} \over \pi} \approx .608173$.

\subsubsection{The six inter-directional correlations all equal  the 
negative of the three intra-directional correlations}

\begin{figure}
\centerline{\psfig{figure=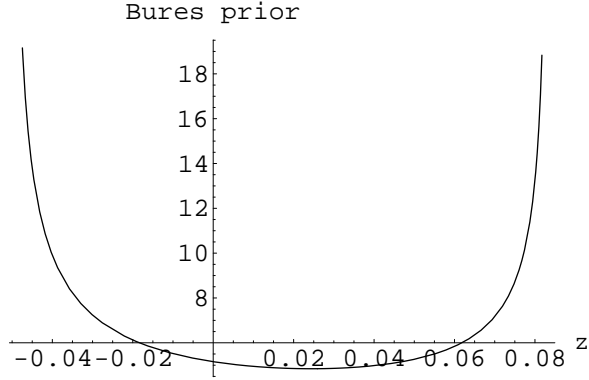}}
\caption{Normalized conditional Bures prior for one-parameter four-level 
scenario 4}
\label{ellipticplot}
\end{figure}

Another one-dimensional scenario of possible interest is one in which 
we set the intra-directional correlations
to $\zeta$ and the inter-directional ones to $-\zeta$. Now, the range of
feasibility is $\zeta \in [-{1 \over 20},{1 \over 12}]$ and the interval of
separability is $\zeta \in [-{1 \over 20},{1 \over 20}]$.
Now, application of the Dittmann formula yields
\begin{equation} \label{gadzooks}
12 { 3 - 20  \zeta \over ( 4 \zeta -1) (12 \zeta - 1) (1 + 20 \zeta)},
\end{equation}
the square root of which gives us the unnormalized Bures prior.
Since the integrations involved yield various elliptic functions, we have to
resort to numerical methods to obtain the Bures probability of separability, 
that is, .702675. In Fig.~\ref{ellipticplot}, 
we plot the associated Bures probability
density function.

\subsubsection{Three scenarios for which the probabilities
 of separability are simply 1}

If we set all nine  (inter- and intra-) directional correlations to 
one value $\zeta$,
and the other six (Stokes/Bloch) 
parameters to zero, so that again the two reduced systems
are fully mixed in nature, then proceeding along the same lines as 
 above, we obtain
the particularly simple conditional  Bures probability distribution,
\begin{equation}
{12 \over \pi (1- 144 \zeta^{2})^{1 \over 2}},
\end{equation}
over the feasible range $\zeta \in [-{1 \over 12}, {1 \over 12}]$.
However, all the states in this range are separable, so the associated 
probability (Bures or otherwise) of separability is simply 1.

If we (formally, but somewhat unnaturally) 
set all fifteen parameters to $\zeta$, say, then the 
conditional Bures prior is
proportional to
\begin{equation}
{2 (3- 20 \zeta)^{1 \over 2} \over (1 +12 \zeta - 336 \zeta^{2}
 + 576 \zeta^{3})^{1 \over 2}}.
\end{equation}
Though version 3 of MATHEMATICA failed (exceeding its iteration limit 
of 4096)
to  integrate over $\zeta \in [- {1 \over 4 (3 + 2 \sqrt{3})} \approx 
-.0386751, {1 \over 12}]$, version 4 (as shown by M. Trott) yielded
\begin{equation}
{1 \over 3} \sqrt{{2 \over 33} (6 + \sqrt{3})} \Pi
\lgroup {5 \over 33} (6 + \sqrt{3}); \sin^{-1}{\lgroup \sqrt{{1 \over 11} (13
 - 4 \sqrt{3})} \rgroup} \vert {1 \over 11} (13 + 4 \sqrt{3}) \rgroup.
\end{equation}
In any case, all the $4 \times 4$ density matrices
in this one-dimensional set are separable, as well. Another scenario in which
the probability of separability is unity, is one in which the three 
intra-directional correlations ($\zeta_{i i}$) 
are all zero, and the two systems are 
{\it anti}-correlated in different directions, that is $\zeta_{i j} = - 
\zeta_{j i}$.
\subsubsection{Rains-Smolin entangled states}
 On p. 182 of \cite{rains}, Rains presents a one-parameter ($x$) set of
$4 \times 4$ density matrices, apparently communicated to him by Smolin.
The corresponding normalized Bures prior for this set of  
states --- defined over the range of feasibility 
$x \in [-u,u], u = \sqrt{807599} / 175 \approx 5.13523$ --- is
\begin{equation}
{175 \over \pi (807599 -30625 x^2)^{1 \over 2}}.
\end{equation}
None of the members of this set is separable.

\subsubsection{Two-qubit Werner states}

\begin{figure} 
\centerline{\psfig{figure=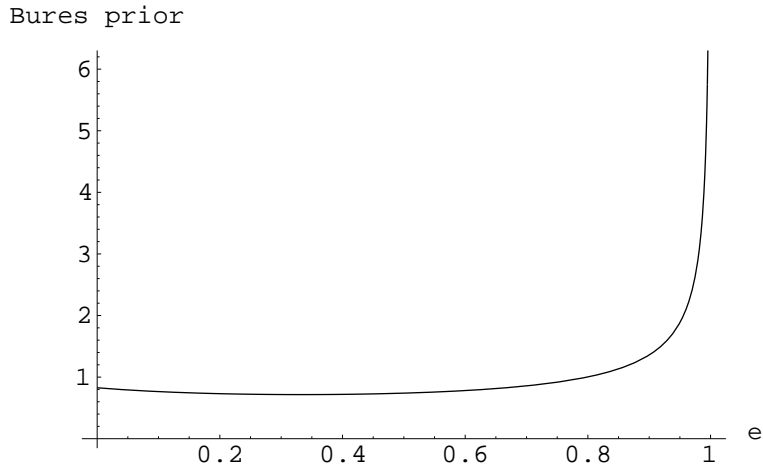}}
\caption{Bures conditional probability distribution (\ref{wq}) over the 
two-qubit Werner states}
\label{qubitWerner}
\end{figure}

It is of some interest that all the Bures 
conditional probabilities of separability we obtained
in the various one-dimensional scenarios above are substantially larger than
the approximate estimate of .1  for the fifteen-dimensional set of 
$4 \times 4$ density 
matrices, obtained on the basis of 
(unfortunately, but perhaps unavoidably, 
rather crude) {\it numerical} methods in
\cite{slatZHSL}. One does, however, obtain a 
(somewhat smaller) probability of separability of
${1 \over 4}$ for the 
two-qubit ``Werner states'' \cite{werner}. These are mixtures of the
fully mixed state and a maximally entangled state, with weights $1-\epsilon$ 
and $\epsilon$, respectively. (In terms of our other set of parameters,
 the three  intra-directional correlations are all equal to
$-\epsilon / 4$, while the remaining twelve parameters are zero.
A referee has suggested that the parameter $\epsilon$ is comparable to 
the {\it visibility} in optics \cite{michelson}, that is $(I_{max} - I_{min}) 
/(I_{max}+I{min})$, where $I$ is the intensity.)
The range of feasibility is $\epsilon \in
[0,1]$ and of separability, $[0,{1 \over 3}]$. The Bures conditional
probability distribution (Fig.~\ref{qubitWerner}) is
\begin{equation} \label{wq}
{3 \sqrt{3} \over \pi (4 + 8 \epsilon -12 \epsilon^{2})^{1 \over 2}}.
\end{equation}

\subsection{{\it Two}-parameter $2 \otimes 2$
systems} \label{wow}

\subsubsection{Two intra-directional correlations are equal and the third 
one, free}

Now we modify scenario 2 of sec.~\ref{scenario2},
 in that we set two intra-directional
correlations again to a common value, call it $\zeta$, and the third, not to
$-\zeta$ this time, but to an independent parameter, call it $\eta$.
(The remaining parameters --- the six Stokes/Bloch ones and the six
 inter-directional correlations stay fixed at zero.)
The normalized conditional Bures prior is then
\begin{equation} \label{2parameter}
{8 \sqrt{2}  \over \pi  {\lgroup (1 + 4 \eta)
 ((1 - 4 \eta)^{2} -64 \zeta^{2}) \rgroup}^{1 \over 2}}.
\end{equation}

\begin{figure}
\centerline{\psfig{figure=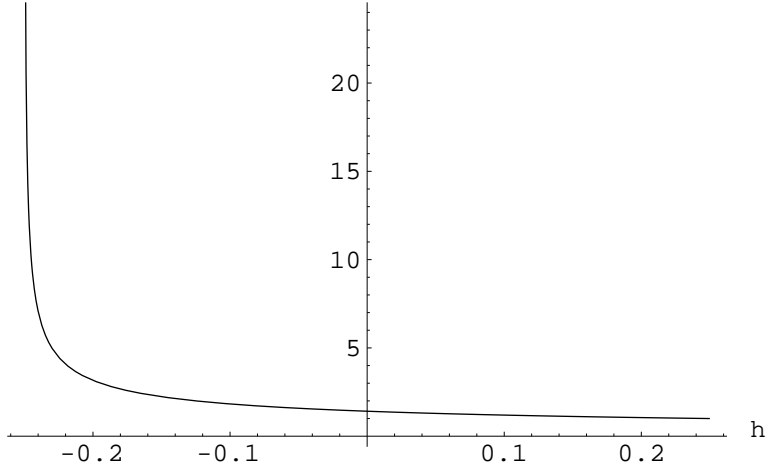}}
\caption{Univariate
 marginal Bures prior probability distribution for two-parameter
four-level scenario}
\label{mb}
\end{figure}

The  (triangular-shaped) 
range of {\it feasibility} over which we integrated to normalize the 
(conditional)
Bures volume element extends in the
$\eta$-direction from 
$-{1 \over 4}$ to ${1 \over 4}$. In this triangle,
 we integrated first over $\zeta$ from
 ${(-1+ 4 \eta) \over 8}$ to ${(1 - 4 \eta) \over 
8}$.
 The part of the (rhombus-shaped) range of {\it separability}
 for  $\eta \in [0,
{1 \over 4}]$ coincides with the feasible domain, and for $\eta \in 
[-{1 \over 4},0]$ extends over   $\zeta \in 
[ -{(1 + 4 \eta) \over 8}, {(1 +  4 \eta ) \over 8}]$.
The univariate 
marginal probability distribution (Fig.~\ref{mb}) 
of (\ref{2parameter}) over $\eta$
is $\sqrt{2} / \sqrt{1 + 4 \eta}$.

The probability of separability for this two-parameter four-level 
scenario is, then, remarkably simply,
$ \sqrt{2} - 1 \approx .414214$ --- being somewhat less than 
the ${1 \over 2}$ of the related scenario 2 of sec.~\ref{scenario2}.
(Of this total figure, $1 - {1 \over \sqrt{2}} \approx .292893$ comes from 
the integration over $\eta > 0$ and ${3 \over \sqrt{2}} - 2 \approx 
.12132$ from the other half of the rhomboidal separability region,  
that is for $\eta < 0$. 
This second result required the use of version 4 of MATHEMATICA, and I thank 
Michael Trott for his assistance.)

\subsubsection{States inferred by the principle of maximum nonadditive 
(Tsallis) 
entropy}

Here the two variables parameterizing the $4 \times 4$ density matrices are
the $q$-expected value ($b_{q}$ --- ``internal energy'') and
 the $q$-variance ($\sigma^{2}_{q}$) of the
Bell--Clauser-Horne-Shimony-Holt (Bell-CHSH) observable 
 \cite{clauser} or ``Hamiltonian'' 
used by Abe and Rajagopal \cite{abe} (cf. \cite{raja}) in their effort 
to avoid  {\it fake} entanglement when only 
$b_{q}$ is employed in the Jaynes maximum entropy 
inference scheme \cite{HoRothree}.
We know from \cite{abe} that the feasible region
 is determined by $0 \leq b_{q} \leq 2 \sqrt{2}$ and
$2 \sqrt{2} b_{q} \leq \sigma^{2}_{q} \leq 8$.

\begin{figure}
\centerline{\psfig{figure=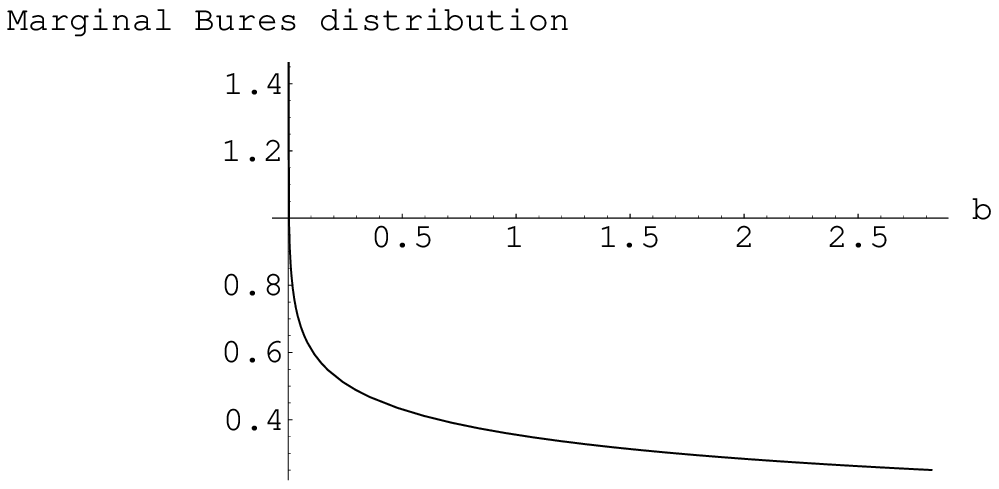}}
\caption{Marginal Bures 
prior probability distribution over the expected value $b_{1}$}
\label{tyu}
\end{figure}

 Let us, first, set the positive parameter $q$ 
indexing the Tsallis entropy to 1.
(``It is of interest to note that for $q > 1$, indicating the subadditive 
feature of the Tsallis entropy, the entangled region is small and enlarges as
ones goes into the superadditive regime, where $q < 1$'' \cite{abe}.) 
Then, the corresponding Bures prior probability distribution --- again 
applying the formula of Dittmann --- is
\begin{equation} \label{kgj}
{1 \over \pi (8 - \sigma_{1}^{2})^{1 \over 2}
  (\sigma_{1}^{2} - 8 b_{1}^{4})^{1 \over 2}}.
\end{equation}
In Fig.~\ref{tyu}, we show the univariate marginal probability distribution
of (\ref{kgj}) --- having integrated it over the parameter 
$\sigma_{1}^{2}$ --- for the expected value $b_{1}$.
The region of {\it separability} is determined \cite[eqs. (11), (12)]{raja} 
by the supplementary requirements
that $\sigma_{1}^{2} \leq 8 - 2 \sqrt{2} b_{1}$ and $b_{1} \leq \sqrt{2}$.
The probability of separability is, then, $\sqrt{2}- 1 \approx 
.414214$ (cf. \cite[Fig. 1(d)]{abe}).

\begin{figure}
\centerline{\psfig{figure=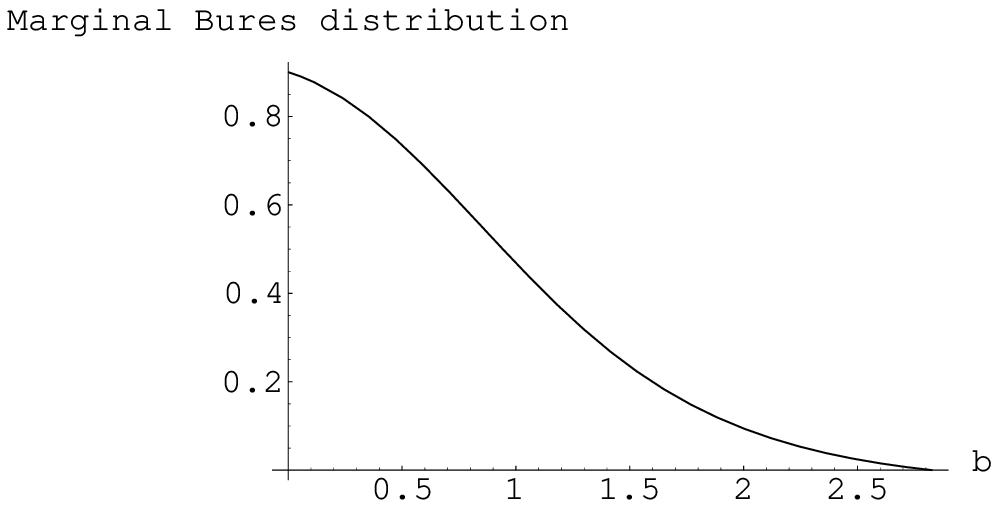}}
\caption{Marginal Bures prior probability distribution over the expected
value $b_{1 \over 2}$}
\label{fas}
\end{figure}

For $q= {1 \over 2}$, the corresponding Bures probability distribution is
(cf. (\ref{kgj}))
\begin{equation} \label{lsw}
{32 \over \pi (32 + 4 {b_{1 \over 2}}^{2} + (\sigma_{1 \over 2}^{2} - 8) 
\sigma_{1 \over 2}^{2})^{3 \over 2}}.
\end{equation}
In Fig.~\ref{fas}, we show the marginal probability distribution of 
(\ref{lsw}) over the $q$-expectation value $b_{1 \over 2}$.
The probability of separability is then  again, quite remarkably,
 $\sqrt{2} - 1 $. (The domain of integration is now 
determined by the supplementary requirements that 
$\sigma_{1 \over 2}^{2} \leq 8 +2 \sqrt{2} b_{1 \over 2} - 
2 \sqrt{2} \sqrt{ b_{1 \over 2} (4 \sqrt{2} + b_{1 \over 2})}$ and
$b_{1 \over 2} \leq 4 - 2 \sqrt{2}$.)
So, it would appear from our two analyses that the probability of separability
is {\it independent} of the particular choice of $q$. (Certainly, whether or 
not this is so bears further investigation. However, we have encountered 
initial computational 
difficulties in obtaining results for other choices of the index
$q$.)

Of course, it would be of interest to consider the index of the Tsallis
entropy $q$ as a third intrinsic variable 
parameterizing the joint states of the two qubits, in addition to $b_{q}$ and
$\sigma_{q}^{2}$, but doing so would appear to  exceed current computational 
capabilities.

\subsection{{\it Three}-parameter 
$2 \otimes 2$ systems} \label{secthreec}

\subsubsection{The three intra-directional correlations are independent} 

If we now modify the scenario immediately above by letting all three 
intra-directional correlations be independent of one another --- while
 maintaining the other 
twelve
(Stokes/Bloch and inter-directional correlation)
 parameters of the four-level systems at zero --- the conditional Bures
prior takes the form (symmetric in the three free parameters)
\begin{equation} \label{hsdc}
{8 \over {\lgroup (-1 + 4 \zeta - 4 \eta - 4 \kappa) (1 + 4 \zeta + 4 \eta
-4 \kappa) (1 + 4 \zeta - 4 \eta + 4 \kappa) (-1 + 4 \zeta + 4 \eta +
4 \kappa) \rgroup}^{1 \over 2}}.
\end{equation}
The normalization factor, by which this must be divided to yield a probability
distribution is ${\pi^{2} \over 8}$.
The Bures probability of separability is ${2 \over \pi} - {1 \over 2} 
\approx .13662$. 

To obtain these results, we employed the
 change-of-variables, 
$\kappa = -{1 \over 4} + \zeta - \eta -{\upsilon \over 4}$, that is,
$\upsilon = -1 + 4 \zeta - 4 \eta - 4 \kappa$.
The integration over the domain of feasibility was obtained using the 
ordered limits, $\zeta \in [-1/4,1/4]$; $\upsilon \in [2 (-1 + 4 \zeta),0]$;
and 
$\eta \in [(-2 - \upsilon)/8,(8 \zeta -\upsilon)/8]$.
For the domain of separability, we employed $\zeta \in [0,1/4]$; $\upsilon
 \in [2 (-1 + 4 \zeta),0]$; and 
$\eta \in [(-2 + 8 \zeta - \upsilon)/8,-\upsilon / 8]$.
The univariate marginal probability distribution of (\ref{hsdc}) over 
$\zeta \in [-1/4,1/4]$ is the uniform one.

\subsubsection{Diagonal density matrices}

As a simple exercise of interest, we have analyzed --- again with the use of 
Dittmann's general formula for the Bures metric tensor --- four-level
 {\it diagonal} density matrices, having 
entries denoted $x,y,z,1-x-y-z$.
(All such density matrices are separable.)
The Bures prior probability distribution over the three-dimensional simplex 
spanned by these entries is, then, 
 simply the Dirichlet distribution
\begin{equation} \label{milton}
{1 \over \pi^{2} (x y z (1-x - y - z))^{1 \over 2}},
\end{equation}
which serves as the prior probability distribution
(``Jeffreys' prior'') based on the Fisher information 
metric for a quadriinomial distribution
\cite{kass,bernardo}.
This result is, thus, consistent with
 the study of Braunstein and Caves \cite{braun},
in which the Bures metric was obtained by maximizing the Fisher information 
 over all quantum measurements, not just ones described by
one-dimensional orthogonal projectors.

\subsection{A {\it Four}-parameter $2 \otimes 2$ system}
 \label{ldcf}

\subsubsection{One-parameter unitary transformation of diagonal density
matrices}

If we transform the (three-parameter) diagonal density matrices discussed 
immediately above by a 
({\it one}-parameter)
$4 \times 4$  unitary matrix,
\begin{equation} \label{firstunitary}
U_{1} = e^{i w P} = \pmatrix{1 & 0 & 0 & 0 \cr 0 & \cos{w} & \sin{w} & 0 \cr
                            0 & -\sin{w} & \cos{w} & 0 \cr 0 & 0 & 0 & 1}
\end{equation}
 where $P$ is a traceless
Hermitian matrix (one of the standard generators of $SU(4)$ 
\cite{stancu}) 
with its only nonzero entries being $i$ in the (3,2) cell and
$-i$ in the (2,3) cell, we find that the (unnormalized) Bures prior is (cf. 
(\ref{milton}))
\begin{equation} \label{pidc}
{\sqrt{(y-z)^2} \over 8 \lgroup x y z (y+z) (1-x-y-z)\rgroup^{1 \over 2}},
\end{equation}
being independent of the fourth (unitary) parameter $w$.
To normalize this prior over the product of the three-dimensional simplex
and the interval $w \in [0, 2 \pi]$, we need to multiply it by
 ${3 \over \pi^{2}}$.
The additional separability requirement is that
$w \in [-u,u]$, where 
\begin{equation}
u ={\sin^{-1} ({2 \sqrt{ x - x^2 - x y - x z} \over \sqrt{y^2 - 2 y z + z^2}}) 
\over 2}.
\end{equation}
The approximate Bures probability of separability is, then, .112, similar 
to the (unrestricted) estimate in \cite{slatZHSL}.

\subsection{{\it One}-parameter $3 \otimes 3$ 
systems --- the two-{\it qutrit} Werner states}

\begin{figure}
\centerline{\psfig{figure=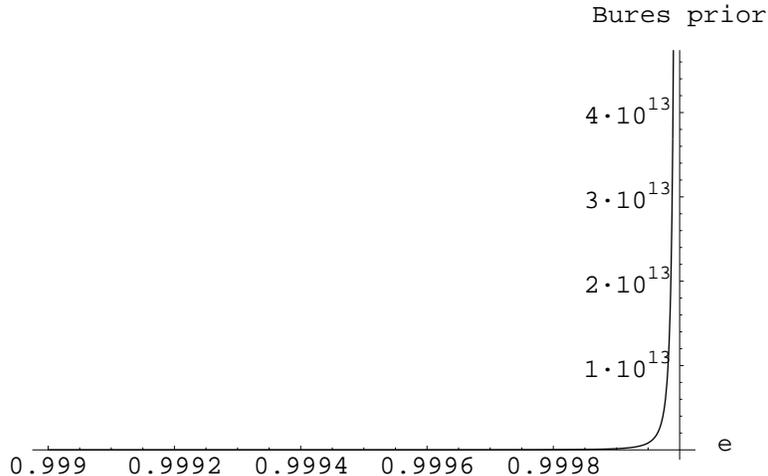}}
\caption{Bures conditional (unnormalized) measure --- that is, 
the square root of the ratio (\ref{qutrit}) --- over the 
two-qutrit Werner states}
\label{qutritWerner}
\end{figure}

Caves and Milburn \cite[sec. III]{cavesmilburn} have constructed
two-{\it qutrit} Werner states. (Such states violate the partial transposition 
criterion for separability, while satisfying a certain 
reduction criterion, the violation of which implies 
distillability \cite{mhph}.) Again applying Proposition 1 of 
Dittman \cite{ditt2} to this one-dimensional set of $9 \times 9$ 
density matrices, we obtained for 
the (unnormalized) conditional Bures prior, the 
{\it square root} of the ratio of 
\begin{equation} \label{qutrit}
-16 (2 +7 \epsilon) (496 + 14384 \epsilon + 179472 \epsilon^2 +
1269568 \epsilon^3 + 5676488 \epsilon^4 +
\end{equation}
\begin{displaymath}
 16753596 \epsilon^5 
+ 31419646 \epsilon^6 + 31863023 \epsilon^7 + 14859999 \epsilon^8) 
\end{displaymath}
to
\begin{displaymath}
 3 (-1+\epsilon)^5 (1+8 \epsilon) (31 + 161 \epsilon)
(31 + 603 \epsilon + 3993 \epsilon^2 + 8981 \epsilon^3).
\end{displaymath}
The range of separability is $[0,{1 \over 4}]$ 
\cite{cavesmilburn}. (If the maximally entangled 
component of the Werner state is replaced by an arbitrary $9 \times 9$ 
density matrix, the range of separability for the resulting mixture 
must include
$ [0,{1 \over 28}] $ \cite{cavesmilburn}.) The integral of the square 
root of the ratio (\ref{qutrit}) over this range is 1.05879, while over 
$[0,.999999]$, it is $9.62137 \times 10^9$. The conditional Bures probability 
of separability of the two-qutrit Werner states, thus, appears to be 
vanishingly small (cf. \cite{slateressent,clifton}. 
The ``maximally entangled states of two qutrits are more 
entangled than maximally entangled states of two qubits'' \cite{cavesmilburn}.
The two-qutrit Bures prior (Fig.~\ref{qutritWerner}) 
is much more steeply rising than the two-qubit 
one displayed in Fig.~\ref{qubitWerner}.
(Note, of course, the difference in the scale employed in the two plots.)

\subsection{{\it One}-parameter $2 \otimes 3$ systems}

It clearly constitutes a  challenging task to extend (cf. sec.~\ref{wow}) 
the series of
one-dimensional analyses in sec.~\ref{secthree}  
to $m$-dimensional ($m > 1$) subsets of the 
fifteen-dimensional set of  $4 \times 4$ density matrices, and
{\it a fortiori} to $n \times n$ density matrices, $n > 4$. Only in highly 
special cases, does it appear that the use of {\it exact}
 integration methods, such 
as exploited above, will succeed,
 and recourse will have to be had to {\it numerical} 
techniques, such as were advanced in \cite{slatZHSL,slateressent}.
We should note, though, that in those two 
studies, numerical {\it integration}
procedures were not readily applicable. They would be available if
one has, as here,
 {\it explicit} forms for the Bures prior, and can suitably define the
limits of integration --- that is, the boundaries of the sets of density 
matrices (and separable density matrices) under analysis.
\subsubsection{Scenario 1}

\begin{figure}
\centerline{\psfig{figure=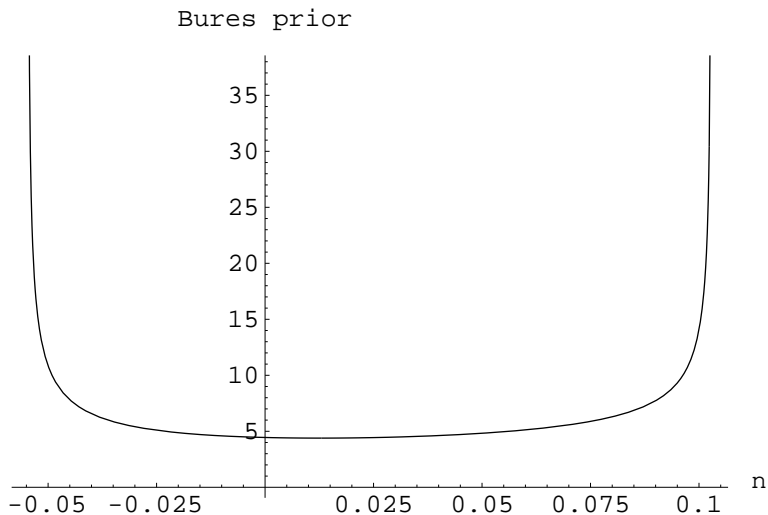}}
\caption{Normalized conditional Bures prior
 over the single parameter ($\nu$)  of the 
{\it six}-level system (\ref{sixlevel})}
\label{3l1}
\end{figure}

As an illustration of the application of numerical integration 
techniques to such a higher-dimensional scenario, we have 
considered the one-parameter ($\nu$) family of $6 \times 6$ density matrices,
\begin{equation} \label{sixlevel}
{ 1 \over 6} \pmatrix{ 1 + 2 \sqrt{3} \nu & 0 & 6 \nu & 0 & 0 & 0 \cr
0 &  1 + 2 \sqrt{3} \nu & 0 & 0 & 0 & - 12 \mbox{i} \nu \cr
6 \nu & 0 &  1 -4 \sqrt{3} \nu & 0 & 0 & 0 \cr
0 & 0 & 0 &  1 - 2 \sqrt{3} \nu & 0 & - 6 \nu \cr
0 & 0 & 0 & 0 &  1 - 2 \sqrt{3} \nu & 0 \cr
0 & 12 \mbox{i} \nu & 0 & -6 \nu & 0 &  1 +4 \sqrt{3} \nu \cr},
\end{equation}
the $2 \times 2$ and $3 \times 3$ reduced systems of which are fully mixed.
The range of feasibility is $\nu \in [-.0546647,.10277]$ and that of 
separability is $\nu \in [-.0546647,.0546647]$. 
(Again, we apply the  
positive partial transposition Peres-Horodecki condition, sufficient as 
well as necessary for both the  $2 \otimes 2$ and $2 \otimes 3$ systems 
\cite{horo3}.) We have determined the 
corresponding (conditional) Bures prior, again based on Proposition 1 
in \cite{ditt2}.
The probability of separability is the ratio of the 
integrals of the prior over 
these two intervals. This turns out to be .607921 --- as seems plausible 
from the plot of the 
(conditional) Bures prior over the feasibility range in Fig.~\ref{3l1}.
This figure displays a minimum at $\nu = 0$ (cf. Fig.~\ref{ellipticplot}).
For $\nu < 0$, we numerically integrate, as well as plot,
 the absolute value of the 
(imaginary) Bures prior in this part of the parameter range.

\subsubsection{The qubit-qutrit Werner states}

\begin{figure}
\centerline{\psfig{figure=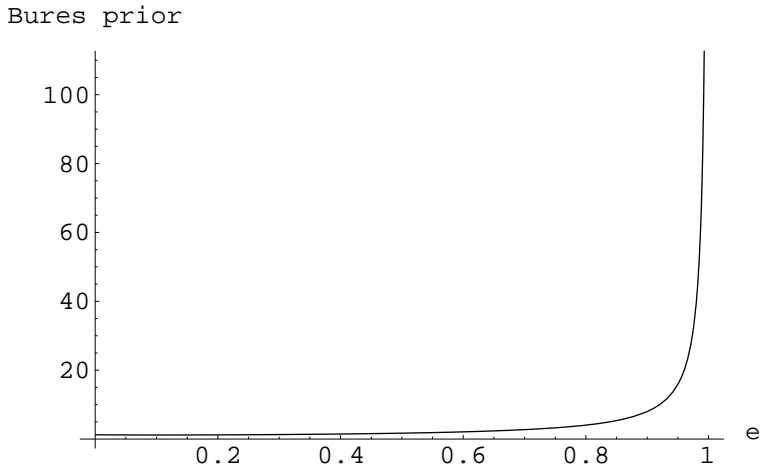}}
\caption{Bures conditional (unnormalized) measure --- that is, the 
square root of the ratio (\ref{bitr}) --- over the 
qubit-qutrit Werner states}
\label{bittrit}
\end{figure}

The probability of 
separability also appears to be vanishingly small for the ``hybrid''
qubit-qutrit Werner states (cf. \cite{slateressent}).  The greatest 
degree of entanglement one can hope to
achieve is to have the qubit in a fully mixed state, and the qutrit in
a degenerate state, with spectrum ${1 \over 2}$, ${1 \over 2}$ and 0.
The range of separability can then be shown to be $[0, {1 \over 4}]$
 \cite{vidal}.
(For the analysis of {\it multi}-qubit Werner states, see \cite{dur}.)
The (unnormalized) conditional Bures prior 
(Fig.~\ref{bittrit}) is the {\it square root} of the
ratio of 
\begin{equation} \label{bitr}
10 (1 + 2 \epsilon) (26 + 286 \epsilon + 1236 \epsilon^{2} +2506 \epsilon^{3}
+2021 \epsilon^{4})
\end{equation}
to
\begin{displaymath}
(-1+\epsilon)^{2} (1 + 5 \epsilon) (13 + 32 \epsilon) (13 + 126 \epsilon
+ 429 \epsilon^{2} + 512 \epsilon^{3}).
\end{displaymath}

\section{Concluding Remarks}

We believe  the main contribution of this study is that it
 emphatically reveals --- through its remarkably simple results of an 
exact nature --- the existence of an intimate
connection between two major (but, heretofore, rather 
distinct)  areas of study in quantum physics. By this is meant the study
of: (1) 
metric structures on quantum systems \cite{petzsudar}; and (2) 
entanglement of quantum systems (cf. \cite{bh}).
 It should be noted, however, that
the several {\it exact} Bures probabilities of separability 
 adduced above are all for certain 
qubit-qubit systems (representable by $4 \times 4$ density matrices).
 It would be of interest to see if 
exact results are obtainable for qubit-qutrit 
(representable by $6 \times 6$ density matrices) 
and even larger-sized systems, 
as well as for qubit-qubit systems parameterized by more than three
variables (the maximum possible being fifteen). Such investigations would,
undoubtedly, demand considerable computational resources and sophistication.
(Let us also direct the reader's attention to our study \cite{slateressent},
in which we employ numerical methods to estimate the Bures probability of
separability of the two-party {\it Gaussian} states, forms of 
continuous variable systems. For this purpose, we
employed an  analogue \cite{dgcz} (cf. \cite{muk}) 
of the Peres-Horodecki criterion
for separability \cite{horo3}.)

We also intend to use the Bures probabilities as weights for the 
{\it entanglement of formation}
\cite{woottersentangl}. By doing so, we should be able to order
different scenarios by the total amount of entanglement they involve. 
As a first example, we have found this figure to be equal to .0441763 for
the first (one-parameter) 
scenario we have analyzed above (sec.~\ref{secthreea}), that of three equal
 intra-directional correlations (and all other parameters fixed at zero).

One of the clearer findings of \cite{slatZHSL} was that the Bures 
(minimal monotone) probability
of separability provides an {\it upper} bound on the related probability 
for any monotone metric. So, it would appear that the results
reported above might also
be interpreted as providing upper bounds on any acceptable measure of the 
probability of separability.

Let us also remark that 
a quite distinct
 direction perhaps worthy of exploration is the use of Bures priors
as densities-of-states or structure functions for thermodynamic 
purposes (cf. \cite{slattherm1,slattherm2,plenio,horotherm,poprohr}). 
(However, our initial efforts to find explicit forms for the
partition functions corresponding to the results above have not succeeded.)

\acknowledgments

I would like to express appreciation to the Institute for Theoretical Physics
for computational support in this research, to Michael Trott of 
Wolfram Research for analyzing a number of the 
symbolic integration problems here with 
version 4 of MATHEMATICA, to K. \.Zyczkowski for his 
continuing encouragement and interest, and to M. J. W. Hall for 
helpful insights into
the properties of the Bures metric, as applied to lower-dimensional subsets of
the $(n^2-1)$-dimensional $n \times n$ density matrices.


\begin{references}
\bibitem{slatZHSL} P. B. Slater, J. Phys. A 32, 5261 (1999).
\bibitem{ditt1} J. Dittmann, Sem. Sophus Lie 3, 73 (1993)
\bibitem{ditt2} J. Dittmann, J. Phys. A 32, 2663 (1999).
\bibitem{hub1} M. H\"ubner, Phys. Lett. A 163, 239 (1992).
\bibitem{hub2} M. H\"ubner, Phys. Lett. A 179, 226 (1993).
\bibitem{braun} S. L. Braunstein and C. M. Caves, Phys. Rev. Lett. 
72, 3439 (1994).
\bibitem{HORO3} M. Horodecki, P. Horodecki, and R. Horodecki,
Phys. Rev. Lett. 80, 5239 (1998).
\bibitem{VEDRAL} V. Vedral, Phys. Lett. A 262, 121 (1999).
\bibitem{slateressent} P. B. Slater, {\it Essentially All Gaussian Two-Party 
Quantum States are a priori Nonclassical but Classically Correlated},
quant-ph/9909062 (to appear in J. Opt. B: Quantum Semiclass. Opt).
\bibitem{clifton} R. Clifton and H. Halverson, Phys. Rev. A 61, 012108 (2000).
\bibitem{bloore} F. J. Bloore, J. Phys. A 9, 2059 (1976).
\bibitem{slatjpa} P. B. Slater, J. Phys. A 29, L271 (1996).
\bibitem{slatjpa2} P. B. Slater, J. Phys. A 29, L601 (1996).
\bibitem{slatjmp1} P. B. Slater, J. Math. Phys. 37, 2682 (1996).
\bibitem{slatcomp} P. B. Slater, Phys. Lett. A 247, 1 (1998).
\bibitem{slathall} P. B. Slater, J. Phys. A 32, 8231 (1999).
\bibitem{ksieee} C. Krattenthaler and P. B. Slater, IEEE Trans. Inform.
Th. 46, 801 (2000).
\bibitem{kass} R. E. Kass, Statist. Sci. 4, 188 (1989).
\bibitem{bernardo} J. M. Bernardo and A. F. M. Smith, {\it Bayesian Theory},
(Wiley, New York,).
\bibitem{frieden} B. R. Frieden, {\it Physics from Fisher Information: A
Unification}, (Cambridge Univ., Cambridge, 1998).
\bibitem{murao} H. Murao, Supercomputer 8, 36 (1991).
\bibitem{krattenthaler} C. Krattenthaler, Sem. Lothar. Combin. 42, B42q (1999).
\bibitem{wootters} W. K. Wootters, Found. Phys. 16,  391 (1986).
\bibitem{wootters2} W. K. Wootters and B. D. Fields, Ann. Phys. 191,
 363 (1989).
\bibitem{boya} L. J. Boya, M. Byrd, M. Mims, and E. C. G. Sudarshan, 
{\it Density Matrices and Geometric Phases for $n$-state Systems}, 
quant-ph/9810084.
\bibitem{byrd1} M. Byrd and E. C. G. Sudarshan, J. Phys. A 31, 9255 (1998).
\bibitem{byrd2} M. Byrd, J. Math. Phys. 39, 6125 (1998).
\bibitem{mallesh} K. S. Mallesh and N. Mukunda, Pramana, 49, 371 (1997).
\bibitem{rowe} D. J. Rowe, B. C. Sanders, and H. de Guise, J. Math. Phys. 
40, 3604 (1999).
\bibitem{slattherm1} P. B. Slater, {\it Volume Elements of Monotone Metrics
on the $n \times n$ Density Matrices as Densities-of-States
for Thermodynamic Purposes. I}, quant-ph/9711010.
\bibitem{slattherm2} P. B. Slater, Phys. Rev. E 61, 6087 (2000).
\bibitem{refhoro} R. Horodecki and M. Horodecki, Phys. Rev. A
54, 1838 (1996).
\bibitem{mk1} V. E. Mkrtchian and V. O. Chaltykian, Opt. Commun.
63, 239 (1987).
\bibitem{fano} U. Fano, Rev. Mod. Phys. 55, 855 (1983).
\bibitem{aravind} P. K. Aravind, Amer. J. Phys. 64, 1143 (1996).
\bibitem{linden} N. Linden, S. Popescu, and A. Sudbery,
Phys. Rev. Lett. 83, 243 (1999).
\bibitem{makhlin} Y. Makhlin, {\it Nonlocal Properties of Two-Qubit Gates and
Mixed States and Optimization of Quantum Computation}, quant-ph/0002045.
\bibitem{kz} M. K\'us and K. \.Zyczkowski, {\it Geometry of Entangled
States}, quant-ph/0006068.
\bibitem{horo3} M. Horodecki, P. Horodecki, and R. Horodecki, Phys. 
Lett. A 223, 1 (1996).
\bibitem{ZHSL} K. \.Zyczkowski, P. Horodecki, A. Sanpera, and M. Lewenstein, 
Phys. Rev. A 58, 883 (1998).
\bibitem{zyczpra} K.  \.Zyczkowski, Phys. Rev. A 60, 3496 (1999).
\bibitem{rains} E. M. Rains, Phys. Rev. A 60, 179 (1999).
\bibitem{werner} R. F. Werner, Phys. Rev. A 40, 4277 (1989).
\bibitem{michelson} A. Michelson, Phil. Mag. (5) 30, 1 (1890).
\bibitem{clauser} J. F. Clauser, M. A. Horne, A. Shimony, and R. A. Holt,
Phys. Rev. Lett. 23, 880 (1969).
\bibitem{abe} S. Abe and A. K. Rajagopal, Phys. Rev. A 60, 3461 (1999).
\bibitem{raja} A. K. Rajagopal, Phys. Rev. A 60, 4338 (1999).
\bibitem{HoRothree} R. Horodecki, M. Horodecki, and P. Horodecki, 
Phys. Rev. A 59, 1799 (1999).
\bibitem{stancu} Fl. Stancu, {\it Group Theory in Subnuclear Physics},
(Clarendon, Oxford, 1996).
\bibitem{cavesmilburn} C. M. Caves and G. J. Milburn, {\it Qutrit
Entanglement}, quant-ph/9910001.
\bibitem{mhph} M. Horodecki and P. Horodecki, Phys. Rev. A 59, 4206 (1999).
\bibitem{vidal} G. Vidal and R. Tarrach, Phys. Rev. A 59, 141 (1999).
\bibitem{dur} W. D\"ur and J. I. Cirac, Phys. Rev. 
A 61, 042314 (2000).
\bibitem{petzsudar} D. Petz and C. Sud\'ar, J. Math. Phys. 
37, 2662 (1996).
\bibitem{bh} D. C. Brody and L. P. Hughston, {\it Geometric Quantum
Mechanics}, quant-ph/9906086.
\bibitem{woottersentangl} W. K. Wootters, Phil. Trans. Roy. Soc. Lond., Ser. 
A 356, 1717 (1998).
\bibitem{dgcz} L. M. Duan, G. Giedke, J. I. Cirac, and P. Zoller,
Phys. Rev. Lett. 84, 4002 (2000).
\bibitem{muk} R. Simon, Phys. Rev. Lett 84, 2726 (2000).
\bibitem{plenio} M. B. Plenio and V. Vedral, Contemp. Phys. 39, 431 (1998).
\bibitem{horotherm} P. Horodecki, R. Horodecki, and M. Horodecki, 
Acta Phys. Slov. 48, 141 (1998).
\bibitem{poprohr} S. Popescu and D. Rohrlich, Phys. Rev. A 56, R3319 (1997).

\end{references}
\end{document}